\documentclass[pre,aps,onecolumn,superscriptaddress,floatfix]{revtex4-1}

\usepackage{amssymb}
\usepackage{epsfig}
\usepackage{amsmath}
\usepackage{subfigure}
\usepackage{graphicx}
\usepackage{textcomp}
\usepackage{url}
\usepackage{float}
\usepackage{color}
\usepackage{cases}
\usepackage[normalem]{ulem}

\usepackage[titletoc,title]{appendix}

\usepackage[colorlinks=true, urlcolor=blue, anchorcolor=blue, citecolor=blue,filecolor=blue,linkcolor=blue,menucolor=blue%,pagecolor=blue
]{hyperref}

\usepackage{color}

\newcommand{\be}{\begin{equation}}
\newcommand{\ee}{\end{equation}}
\newcommand{\bea}{\begin{eqnarray}}
\newcommand{\eea}{\end{eqnarray}}

\begin{document}

\title{Fractional Brownian motion in confining potentials: non-equilibrium distribution tails and optimal fluctuations}

\author{Baruch Meerson}
\email{meerson@mail.huji.ac.il}
\affiliation{Racah Institute of Physics, Hebrew University of
Jerusalem, Jerusalem 91904, Israel}

\author{Pavel V. Sasorov}
\email{pavel.sasorov@gmail.com}
\affiliation{ELI Beamlines Facility, Extreme Light Infrastructure ERIC, 252 41 Dolni Brezany,  Czech Republic}

\begin{abstract}
At long times, a fractional Brownian particle in a confining external potential reaches a non-equilibrium (non-Boltzmann)
steady state. Here we consider scale-invariant power-law potentials $V(x)\sim |x|^m$, where $m>0$, and employ the optimal fluctuation method (OFM) to determine the large-$|x|$ tails of the steady-state probability distribution $\mathcal{P}(x)$ of the particle position. The calculations involve finding the optimal (that is, the most likely) path of the particle, which determines these tails, via a minimization of the exact action functional for this system, which has recently become available.  Exploiting dynamical scale invariance of the model in conjunction with the OFM ansatz, we establish the large-$|x|$ tails of $\ln \mathcal{P}(x)$ up to a dimensionless factor $\alpha(H,m)$, where  $0<H<1$ is the Hurst exponent. We determine $\alpha(H,m)$ analytically (i) in the limits of $H\to 0$ and $H\to 1$, and (ii) for $m=2$ and arbitrary $H$, corresponding to the fractional Ornstein-Uhlenbeck (fOU) process. Our results for the fOU process are in agreement with the previously known exact $\mathcal{P}(x)$ and autocovariance. The form of the tails of $\mathcal{P}(x)$ yields exact conditions,  in terms of $H$ and $m$, for the particle confinement in the potential.
For $H\neq 1/2$, the tails encode the non-equilibrium character of the steady state distribution, and we observe violation of time reversibility of the system except for $m=2$. To compute the optimal paths and the factor $\alpha(H,m)$  for arbitrary  permissible $H$ and $m$, one needs to  solve an (in general nonlinear) integro-differential equation.  To this end we develop a specialized numerical iteration algorithm which
accounts  analytically for an intrinsic cusp singularity of the optimal paths for $H<1/2$.

\end{abstract}
\maketitle

\section{Introduction}

The Mandelbrot -– van Ness fractional Brownian motion (fBm) \cite{Kolmogorov,Mandelbrot,Qian2003}, a Gaussian
non-Markovian process with stationary increments, provides a prototypical
example of a system with long-range temporal correlations  and anomalous diffusion.
Here we focus on the following question: how are the properties of the fBm modified in the presence
of confinement? A natural model system for studying a confined fBm
involves overdamped stochastic
motion of a fractional Brownian particle in an external potential  \cite{Metzler2021,Metzler2022,MBO,Metzler2023}. Here we consider a family of scale-invariant power-law potentials
$V(x) = k |x|^m/m$, where $m>0$. This brings about a non-Markovian Langevin equation
\begin{equation}\label{Langevin1}
\dot{x} = f[x(t)]+\xi(t)
\end{equation}
with the deterministic force
\begin{equation}\label{f(x)}
f(x) = - \frac{dV(x)}{dx} = -k |x|^{m-1} \,\text{sign}\,x
\end{equation}
and the long-range correlated fractional Gaussian noise (fGn) $\xi(t)$. In the absence of external potential, the particle position $x(t)$ in Eq.~(\ref{Langevin1})
follows the fractional Brownian motion (fBm),  introduced by Kolmogorov \cite{Kolmogorov} and by Mandelbrot
and van Ness (MvN) \cite{Mandelbrot}. Like the standard Brownian motion, the MvN fBm is a scale-invariant Gaussian process. It is defined, for all $|t|<\infty$, by its mean (which can be set to zero) and the autocovariance \cite{Mandelbrot}
\begin{equation}\label{kappafBm}
\!\kappa(t,t')=\langle x(t) x(t')\rangle= D \left(|t|^{2H}\!+\!|t'|^{2H}\!-\!|t-t'|^{2H}\right),
\end{equation}
where $0<H<1$ is the Hurst exponent, and $D$ is the generalized diffusion constant with units $\text{length}^2/\text{time}^{2H}$. The standard Brownian motion corresponds to $H=1/2$. For $H\neq 1/2$, the MvN process is  non-Markovian and long-range correlated. It describes subdiffusion and superdiffusion for $H<1/2$ and $H>1/2$, respectively. For $H=1/2$ the standard Brownian motion is recovered.

The fGn -- the time derivative of the MvN fBm -- is a stationary Gaussian process, and it is also  scale invariant. Its autocovariance $c(t_1-t_2)=\langle \xi(t_1) \xi(t_2)\rangle$
is the following \cite{Mandelbrot,Qian2003}:
\begin{equation}
c(\tau)
 = 2DH\frac{d}{d\tau}\left(|\tau|^{2H-1}\text{sign}\,\tau\right)
 \equiv D \frac{d^2}{d\tau^2}|\tau|^{2H}\,.
 \label{corry}
\end{equation}
In the particular case of $H=1/2$, Eq.~\eqref{corry} yields
\begin{equation}\label{white}
c(\tau) = D\frac{d}{d\tau} \,\text{sign}\,\tau=2D\delta(\tau)\,,
\end{equation}
the familiar expression for the white noise.

In the presence of the confining potential $V(x)$, the fractional Brownian particle, as described by Eq.~(\ref{Langevin1}), reaches a
steady state. Unless $H=1/2$, this is a \emph{non-equilibrium} steady state. In particular, the  steady-state probability distribution $\mathcal{P}(X)$ of observing the particle at the position $x=X$ is \emph{not} given by the  Boltzmann formula \cite{Metzler2021}. This and other nonequilibrium features, that we discuss here, make this model interesting not only in particular applications (see Ref. \cite{Metzler2021} for an extensive list of such), but also in the more general context of statistical mechanics far from equilibrium.

For $H\neq 1/2$, the steady-state distribution $\mathcal{P}(X)$ is known exactly  only in the special case of the quadratic potential, $m=2$ \cite{Metzler2021}. Here we focus on the large-$|X|$ tails of $\mathcal{P}(X)$ for all $H$ and $m$ that allow for the particle confinement. These (identical) tails describe large deviations of the particle position from its mean value $X=0$.   Exploiting the corresponding large parameter,  we apply the optimal fluctuation  method, which relies on finding the optimal (that is, the most likely) path of the particle which determines the tails. The calculations involve a minimization of the proper action functional for Eq.~(\ref{Langevin1}), recently obtained in Ref. \cite{MBO}.

Using scale invariance properties of the model in conjunction with the OFM, one can immediately establish the large-$|X|$ asymptotic of the logarithm of $\mathcal{P}(X)$ up to an unknown dimensionless factor $\alpha(H,m)$. We find $\alpha(H,m)$ analytically (i) in the limits of $H\to 0$ and $H\to 1$, and (ii) for $m=2$ and arbitrary $H$. The latter result agrees with the previously known exact expression for $\mathcal{P}(x)$ for $m=2$ \cite{Metzler2021}. The knowledge of the tail asymptotic provides exact conditions for the confinement of the particle  in terms of $H$ and $m$, which agree with previous numerical results \cite{Metzler2022}. As an additional manifestation of the nonequilibrium nature of the steady state,  we also observe violation of the time reversibility of the systems' paths, except for $H=1/2$ or $m=2$.

The minimization of the action functional \cite{MBO} can be reduced to solving an (in general nonlinear) integro-differential equation for the optimal path conditioned on reaching a specified point $x=X$.  We developed a specialized numerical iteration algorithm for this purpose. The algorithm accounts analytically for the intrinsic cusp singularities of the optimal paths  and, in principle, makes it possible to compute the dimensionless factor $\alpha(H,m)$  for any permissible $H$ and $m$.

Here is a plan of the rest of the paper. In Sec. \ref{OFM} we revisit and slightly extend the path integral formulation \cite{MBO} for Eq.~(\ref{Langevin1}), introduce the OFM,
establish the scaling behavior of the distribution tails, obtain the confinement condition, exactly solve the OFM problem for the quadratic potential $m=2$, determine the factor $\alpha(H,M)$ in the limits of $H\to 0$ and $H\to 1$ for any permissible $m$ and
demonstrate the violation of the time reversibility of the system's paths for $H\neq 1/2$.  In Sec. \ref{numerics} we present our numerical algorithm for computing the optimal paths
and evaluating the function $\alpha(H,m)$. A brief summary of main results is presented in Sec. \ref{discussion}.  Some technical details of the numerical algorithm are relegated to the two appendices.

\section{Optimal Fluctuation Method}
\label{OFM}

\subsection{Path integral}

Here we will exploit  exact path-integral representations
for Eq.~(\ref{Langevin1}) which have been recently derived separately for the subdiffusion, $0<H<1/2$, and the superdiffusion, $1/2<H<1$ \cite{MBO,half}. For the subdiffusion, the  action functional of the path integral reads
\begin{equation}\label{Ssub}
\mathcal{S}[x(t)] =\frac{1}{2} \int_{-\infty}^{\infty} dt_1 \int_{-\infty}^{\infty} dt_2\,
  C(t_1-t_2) [\dot{x}-f(x)]_{t_1} [\dot{x}-f(x)]_{t_2} \,,
\end{equation}
where the subscripts $t_1$ and $t_2$ denote the arguments of the functions inside the square brackets. In its turn,
\begin{equation}\label{solC2}
C(\tau) = \frac{\text{cot}\, (\pi H)}{4\pi D H} \frac{1}{|\tau|^{2H}}
\end{equation}
is the inverse kernel,
\begin{equation}\label{inversekerneldef}
\int_{-\infty}^{\infty} dt''\,C(t-t'') c(t'-t'') = \delta(t-t')\,,
\end{equation}
for the autocovariance $c(\tau)$ of the fGn, see Eq.~(\ref{corry}).

For the superdiffusion Ref. \cite{MBO} gives the action functional
\begin{equation}\label{actionforcesuper}
\mathcal{S}[x(t)] = \frac{1}{2} \int_{-\infty}^{\infty} dt_1 \int_{-\infty}^{\infty} dt_2\,Q(t_1-t_2)  \left[\ddot{x}-f'[x(t)]\,\dot{x}(t)\right]_{t_1} \left[\ddot{x}(t)-f'[x(t)]\,\dot{x}(t)\right]_{t_2},
\end{equation}
where
\begin{equation}\label{kernelmore}
Q(\tau) =   \frac{{\rm cot}(\pi H)}{8 \pi D H(1-H) (2H-1)} |\tau|^{2-2H}\,,
\end{equation}
and $f'(x) \equiv df/dx$.

In fact, both expressions (\ref{Ssub}) and (\ref{actionforcesuper}) can be represented as Eq.~(\ref{Ssub}) with the universal kernel
\begin{equation}\label{kernelgeneral}
C(t_1-t_2) = \frac{{\rm cot}(\pi H)\,\partial_{t_1}\!\partial_{t_2}|t_1-t_2|^{2-2H}}{8 \pi D H(1-H) (2H-1)}\,.
\end{equation}
For $H<1/2$ the differentiations in Eq.~(\ref{kernelgeneral}) involve classical functions. For $H>1/2$, the differentiations should be interpreted
in terms of generalized functions, or Schwartz distributions.

It is useful to also present the governing equations in the frequency domain. The action for a given realization of the fGn $\xi(t)$ can be written as
\begin{equation}\label{P250}
S[\xi(t)]=\frac{1}{4\pi}\int\limits_{-\infty}^\infty C_{\omega} \xi_{\omega} \xi_{\omega}^*\, d\omega\, ,
\end{equation}
where  $\xi_{\omega} = \int_{-\infty}^{\infty} dt\,\xi(t) e^{i\omega t}$ is the Fourier transform of the fGn,
\begin{equation}
\label{Comega}
C_{\omega} = \int_{-\infty}^{\infty}dt\, C(t) e^{i\omega t}= \frac{| \omega | ^{2 H-1}}{2D \sin (\pi  H) \Gamma (2 H+1)}
\end{equation}
is the Fourier transform of the inverse kernel $C(t)$, and $\Gamma(\dots)$ is the gamma function. The expression~(\ref{P250}) for the action with this positive $C_\omega$ is well defined for any $0<H<1$.

The Fourier transform of the autocovariance itself is
\begin{equation}
\label{comega}
c_{\omega} = \int_{-\infty}^{\infty}dt\, c(t) e^{i\omega t}=2 D\sin (\pi  H)\Gamma (2 H+1)|\omega| ^{1 -2H}  \,,
\end{equation}
as Eq.~(\ref{inversekerneldef}) becomes simply $c_\omega\, C_\omega=1$.
As a result, we can also express in the frequency domain the action~(\ref{Ssub}):
\begin{equation}
\label{FSsub}
  \mathcal{S}[x(t)] = \frac{1}{4\pi} \int_{-\infty}^{\infty} d\omega \, C_\omega
    [\dot{x}-f(x)]_{\omega}\,
   [\dot{x}-f(x)]_{\omega}^* \,.
\end{equation}

When $H\to 1/2$, the action functional (\ref{Ssub}) reduces to the familiar local action for the white-noise driving:
\begin{equation}\label{Swhite}
 \mathcal{S}[x(t)] = \frac{1}{4D}\int_{-\infty}^{\infty} dt \left[\dot{x}(t)-f(x(t))\right]^2\,.
\end{equation}

\subsection{OFM formulation}

Sufficiently far in the tail,  $\mathcal{P}(X)$ is dominated
by a single \emph{optimal path}, for which we will keep the notation $x(t)$. The optimal path minimizes the action (\ref{Ssub})  subject to the conditions
\begin{equation}\label{X}
x(t=0) = X
\end{equation}
and
\begin{equation}\label{BCs}
x(t\to -\infty)=x(t\to\infty) =0.
\end{equation}
It is convenient to incorporate the condition (\ref{X}) by introducing the modified action
\begin{equation}\label{modified}
\mathcal{S}_{\lambda}[x(t)] = \mathcal{S}[x(t)]-\lambda \int_{-\infty}^{\infty} dt \,x(t) \delta(t)\,.
\end{equation}
where $\mathcal{S}[x(t)]$ is given by Eq.~(\ref{Ssub}). The Lagrange multiplier $\lambda$ should be ultimately expressed through $X$ and the rest of the parameters of the problem.

Now we demand that the linear variation of the modified action (\ref{modified}) vanish. To get rid of the time derivatives of the variations $\delta x(t)$ and $\delta x(t')$ we perform two integrations by parts: with respect to $t$ and $t'$, respectively. Then,  using  the symmetry condition $C(t-t')=C(t'-t)$,  we arrive at the following equation:
\begin{equation}
  \int_{-\infty}^{\infty} dt' C (t-t')\left[f'(x(t))+\partial_{t'}\right] \left[\dot{x}-f(x)\right]_{t'} =-\lambda \delta(t)
\label{inteq1}
\end{equation}
which, along with the conditions~(\ref{X}) and (\ref{BCs}), describe the optimal path $x(t)$. In general, the integro-differential equation~(\ref{inteq1})  is nonlinear. It is linear, however, for the quadratic
potential $m=2$, which will be considered in Sec. \ref{quadratic}.

Once the optimal path $x(t)$ is determined, one can evaluate, up to an unknown pre-exponential factor, the position distribution $\mathcal{P}(X)$ by using the approximate relation $-\ln \mathcal{P}(X) \simeq S(X)$, where $S(X)$ is the action (\ref{Ssub}) evaluated along the optimal path. A useful shortcut in the calculation of the action is provided by the relation
\begin{equation}\label{shortcut}
 \frac{dS}{dX}=\lambda(X)\,,
\end{equation}
which follows from the fact that $X$ and $\lambda$ are conjugate variables,
see, e.g. Ref. \cite{Cunden}. Equation~(\ref{shortcut}) enables one to immediately calculate the action once the Lagrange multiplier $\lambda$ is expressed through $X$, or vice versa.

\subsection{Scaling behavior}
\label{scalingbehavior}
The scale invariance properties of the fGn and of the power-law potential $V(x)=k|x|^m/m$ make it possible to immediately establish some important scaling properties of the action $S(X)$ by using a simple dimensional argument \cite{Barenblatt}.  The dimensionless action $S(X)\simeq -\ln \mathcal{P}(X)$ can depend only on the dimensional parameters $X$, $k$ and $D$ and on the dimensionless parameters $H$ and $m$. The units of $D$ are $\text{length}^2/ \text{time}^{2H}$. The units of $k$ are $\text{length}^{m-1}/\text{time}$, see Eqs.~(\ref{Langevin1}) and (\ref{f(x)}). By virtue of the $\Pi$-theorem due to Buckingham (see, \textit{e.g.} Ref. \cite{Barenblatt}),  there is  only \emph{one} independent dimensionless combination of  $X$, $k$ and $D$. This fact, combined with the $1/D$ scaling of $S$, predicted by Eqs.~(\ref{Ssub}) and~(\ref{kernelgeneral}), leads to the insightful scaling form
\begin{equation}\label{P160}
S(X)= \alpha(H,m)\,\frac{k^{2H} |X|^{2(1-2H+Hm)}}{D}\,.
\end{equation}
Equation~(\ref{P160}) establishes the form of the distribution tails $\mathcal{P}(|X| \to \infty)$ as a function of $|X|$ which is manifestly non-Boltzmann for all $H\neq 1/2$. At this stage, the important dimensionless factor $\alpha(H,m)$ is yet unknown except for $H=1/2$, where $S(X)$ describes the Boltzmann distribution
\begin{equation}\label{Boltzmann}
S_{H=1/2}(X) = \frac{k |X|^{m}}{m D}\,,
\end{equation}
so that $\alpha(1/2,m)=1/m$.

Equation~(\ref{P160}) also enables us to establish the confinement condition which makes the steady state possible. Indeed, for the subdiffusion $0<H<1/2$ the power of $X$ in Eq.~(\ref{P160}) is positive for any $m>0$, automatically implying confinement. For the superdiffusion $1/2<H<1$  the power of $X$  becomes negative if
\begin{equation}\label{P162}
m< m_0(H)\equiv \frac{2H-1}{H}\,.
\end{equation}
In this case the potential $V(x)$ is non-confining, and a steady state is impossible. Therefore, the potential is confining, and the steady state can be reached, for $H<1/2$ (for all $m>0$) or for $H>1/2$ and $m>m_0(H)$. These conditions coincide with those presented in Ref. \cite{Metzler2022}, where they were conjectured by using an analogy with the L\'{e}vy flights, and then verified in numerical simulations. The criterion (\ref{P162}) is most stringent when $H\to 1$.

In the rest of this section we determine $\alpha(H,m)$ exactly for $m=2$, and
find the leading-order asymptotics of $\alpha(H,m)$ in the limit of $H\to 0$ and $H\to 1$.
Before that, however, let us briefly review the simple case of white-noise driving, $H=1/2$.

\subsection{$H=1/2$}
\label{mark}

This case corresponds, for any $m>0$, to thermal equilibrium, and there is no need to address the dynamics in order to find the exact (Boltzmann) distribution. It is still interesting, however, to determine the large-$X$ optimal paths of the particle for different $m$.  This is easily done, because the activation path in equilibrium coincides with the time-reversed relaxation path. The latter can be found, for any $m>0$, by solving the deterministic equation
$\dot{x} = -k |x|^{m-1}\,\text{sign}(x)$ with the initial condition $x(t=0)=X$. As a result, the whole optimal path is
\begin{equation}\label{H05a}
x(t)=\left[X^{2-m}+k (m-2) |t|\right]^{-\frac{1}{m-2}}\,,
\end{equation}
where we assumed that $X>0$. For $m\geq 2$ the optimal paths live on the whole line $|t|<\infty$. Note that as $m>2$ increases the large-$|t|$ power-law tails become very long.

For $m<2$ the optimal paths have a compact support, $|t|<T<\infty$. For example, for $m=1$ one obtains
\begin{equation}\label{m05b}
  x(t) =
    \begin{cases}
        X-k|t|\,, & |t|<\frac{X}{k}\,,\\
       0\,, & |t|>\frac{X}{k}\,.
    \end{cases}
\end{equation}
For a quadratic potential $m=2$, and only in this case, the tails are exponential. In  the next subsection we will
consider this important particular case for arbitrary $0<H<1$.

\subsection{Quadratic potential}
\label{quadratic}

\subsubsection{$m=2$: OFM solution}
\label{m2gen}

For $m=2$ the Langevin equation (\ref{Langevin1}) becomes linear,
\begin{equation}\label{fOU}
\dot{x} = -kx +\xi(t)\,,
\end{equation}
and it describes the fractional Ornstein-Uhlenbeck (fOU) process \cite{Cheredito2003,Kaarakka2015}, a useful generalization of the celebrated Ornstein-Uhlenbeck process \cite{OU} (the latter arises in the particular case of $H=1/2$).

As this linear equation is driven by a Gaussian (albeit non-white) noise, the steady-state distribution $\mathcal{P}$ is also
Gaussian, and it has been known for some time. Here, once the factor $\alpha(H,2)$ is determined, the Gaussian tail
\begin{equation}\label{Sm2}
S_{m=2}(X)= \alpha(H,2)\,\frac{k^{2H} X^{2}}{D}\,,
\end{equation}
predicted by the OFM scaling relation~(\ref{P160}) actually gives the whole exact distribution (upon normalization to unity). Although this factor
is known from the exact solution, see Ref. \cite{Metzler2021}, it is instructive to rederive it within the framework of the OFM. The OFM also provides the (observable and quite informative) optimal path of the system, not easily accessible otherwise.  In addition, as we see below, it provides  the exact autocovariance of the stationary fOU process.

For $m=2$, Eq.~(\ref{inteq1}) for the optimal path $x(t)$ simplifies to
\begin{equation}
  \int_{-\infty}^{\infty} dt' C (t-t')\left[\ddot{x}(t')-k^2 x(t')\right] =-\lambda \delta(t)\,.
\label{inteqlin}
\end{equation}
This linear equation can be easily solved by Fourier transform.  One obtains an algebraic equation,
\begin{equation}\label{algebraic}
 (\omega^2+k^2) C_{\omega} x_{\omega} =  \lambda\,,
\end{equation}
where  $x_{\omega}= \int_{-\infty}^{\infty} dt\,x(t) e^{i\omega t}$.
Equation~(\ref{algebraic}) yields
\begin{equation}\label{xomega}
 x_{\omega} = \frac{\lambda}{(\omega ^2+k^2)C_\omega}=\frac{\lambda c_\omega}{(\omega ^2+k^2)}
  = \frac{2 D \lambda  \sin (\pi  H) \Gamma (2 H+1) |\omega| ^{1-2H}}{\omega ^2+k^2}\,.
\end{equation}
Now we perform the inverse transform, $x(t)=(2\pi)^{-1} \int_{-\infty}^{\infty}d\omega\, x_{\omega} e^{-i\omega t}$.
Demanding that $x(t=0)=X$, we obtain
\begin{equation}\label{lambdavsX}
 \lambda=\lambda(X)=\frac{k^{2 H}X}{D \Gamma (2 H+1)}\,,
\end{equation}
and the optimal path is given, for all $0<H<1$, by
\begin{equation}
\label{xquad1}
\frac{x(t)}{X} = \cosh (k t)-\frac{(k|t|)^{2 H}} {\Gamma(2H+1)}  \,
   _1F_2\left(1;H+\frac{1}{2},H+1;\frac{k^2 t^2}{4}\right)\,,
   \end{equation}
where $_1F_2$ is the hypergeometric function \cite{Wolfram}. Remarkably, Eq.~(\ref{xquad1}) coincides, up to a constant factor, with the known autocovariance $\kappa(t)$ of the \emph{stationary} fOU process \cite{Cheredito2003}. We explain this coincidence in subsubsection \ref{steady}.

\begin{figure}[ht]
  %\centering
  \includegraphics[width=0.45\textwidth]{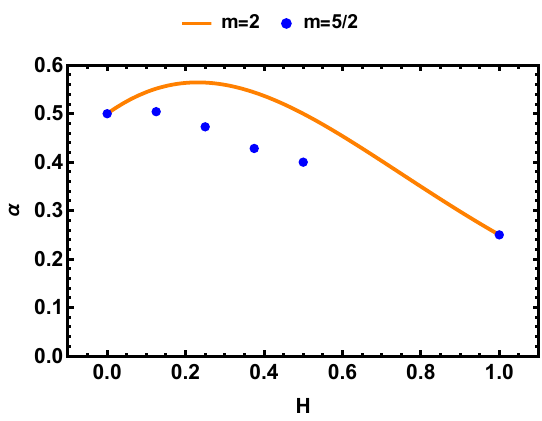}
  \caption{Shown is the coefficient $\alpha(H,m)$ in Eq.~(\ref{P160}) for $m=2$ (orange curve) and $m=5/2$ (blue points). The orange curve corresponds to Eq.~(\ref{alpham=2}). The blue points for $H=0$,1/2 and 1 are plotted in accordance to Eq.~(\ref{threelimits}), whereas the three points on the interval $0<H<1/2$, for $H=1/8$, $1/4$ and $3/8$, are plotted using numerical solutions for the optimal paths, described in Sec.~\ref{numnum}, Eq.~(\ref{alphax(0)}).}
  \label{figa}
\end{figure}

Using Eqs.~(\ref{shortcut}) and (\ref{lambdavsX}), we determine the action:
\begin{equation}\label{actionm=2}
S_{m=2}(H)=\int_0^{X} dX\, \lambda(X) = \frac{k^{2 H} X^2}{2 D \Gamma (2 H+1)}\,.
\end{equation}
Comparing Eqs.~(\ref{Sm2}) and (\ref{actionm=2}), we obtain
\begin{equation}\label{alpham=2}
\alpha(H,2) = \frac{1}{2 \Gamma(2H+1)}\,.
\end{equation}
Interestingly, $\alpha(H,2)$ is a non-monotonic function of $H$, see Fig.~\ref{figa}: it has a maximum $0.5645\dots$ at $H=0.2308\dots$, the only root of the digamma function $\psi ^{(0)}(2 H+1)$ for positive $H$.

Equation~(\ref{actionm=2}) yields the variance of $X$,
\begin{equation}\label{variancem=2}
\text{Var}_{X,m=2} = \frac{D \Gamma(2H+1)}{k^{2H}}\,,
\end{equation}
which perfectly agrees with the exact expression presented in Ref. \cite{Metzler2021}, see also the footnote \cite{half}.

It is useful for the following to rederive these results in the frequency domain.  Using Eq.~(\ref{FSsub}), we obtain
\begin{equation}
\label{AA006}
  \mathcal{S}[x(t)] = \frac{1}{4\pi} \int_{-\infty}^{\infty} d\omega \, C_\omega
   (\omega^2+k^2)x_{\omega}\,
   x_{\omega}^*\,.
\end{equation}
The modified action~(\ref{modified}) becomes
\begin{equation}
\label{AA010}
  \mathcal{S}_\lambda[x(t)] = \frac{1}{4\pi} \int_{-\infty}^{\infty} d\omega \, \left[C_\omega
   (\omega^2+k^2)x_{\omega}\,
   x_{\omega}^* -2\lambda x_\omega\right]\,.
\end{equation}
Its minimization yields  Eq.~(\ref{algebraic}) and then Eq.~(\ref{xomega}). Substituting Eq.~(\ref{xomega}) into Eq.~(\ref{AA006}), we obtain the action
\begin{equation}
\label{AA020}
  \mathcal{S}[x(t)] = \frac{\lambda}{4\pi} \int_{-\infty}^{\infty} d\omega \,
   x_{\omega}\, .
\end{equation}
Hence
\begin{equation}
\label{AA030}
  \mathcal{S} = \frac{1}{2} \lambda X(\lambda)=\frac{1}{2} \lambda(X) X\, ,
\end{equation}
where $X(\lambda)=(2\pi)^{-1}\int_{-\infty}^\infty d\omega\, x_\omega=D\Gamma(2H+1)k^{-2H}$.
Using these expressions, we reproduce Eqs.~(\ref{lambdavsX})-(\ref{alpham=2}). The most important for a further use are Eqs.~(\ref{xomega}) and~(\ref{AA020}).

%%%%%%%%%%%%%%%%%%

%%%%%%%%%%%%%%%%%%

\subsubsection{$m=2$: asymptotics of the optimal paths}
Now let us return to the optimal paths (\ref{xquad1}). Their asymptotic  at $t \to 0$,
\begin{equation}\label{smallt}
  \frac{x(t)}{X}\Bigr|_{k|t|\ll 1}\simeq 1-\frac{(k| t|) ^{2 H}}{\Gamma (2 H+1)}\,,
\end{equation}
exhibits a cusp singularity for $H<1/2$. At $H=1/2$ this singularity becomes a corner singularity, whereas
for $H>1/2$ the optimal path is already once differentiable at $t=0$.

No less instructive is the power-law asymptotic behavior at large $k|t|$,
\begin{equation}\label{larget}
  \frac{x(t)}{X}\Bigr|_{k|t|\gg 1}\simeq \frac{(k|t|)^{2H-2}}{\Gamma(2H-1)}\,,
\end{equation}
which holds at any $H\neq 1/2$. As one can see, the $k|t|\to \infty$ tails become very fat as $H$ approaches $1$. Remarkably, in order to reach the point $x=X>0$ along the least-action path, a subdiffusive particle starting at $x=0$ first moves to the left, and only then to the right.

At $H=1/2$ the optimal path~(\ref{xquad1}) has a simple form $x(t)= X\,e^{-k|t|}$. The same result follows, in the limit of $m\to 2$, from Eq.~(\ref{H05a}).  As we have already mentioned, in this case the activation trajectory $0\to X$ coincides with the time-reversed relaxation trajectory $X\to 0$.

\begin{figure}[ht]
  %\centering
  \includegraphics[width=0.47\textwidth]{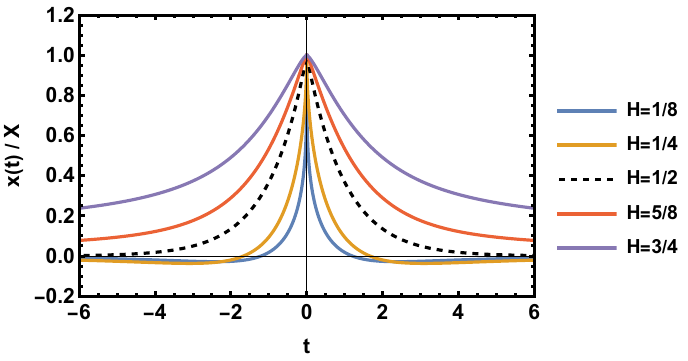}~~~~~
  \includegraphics[width=0.47\textwidth]{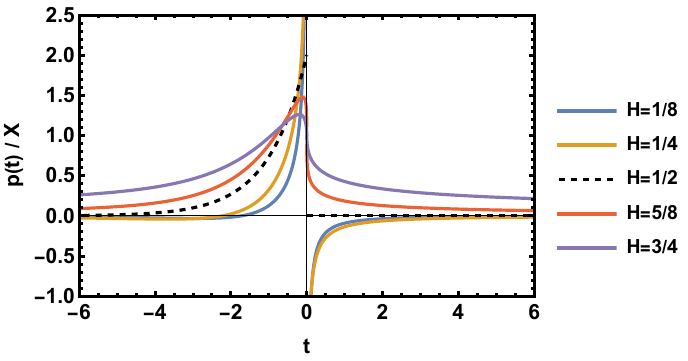}
  \caption{Left panel: The optimal paths $x(t)$ for the quadratic potential $V(x)=x^2/2$ for several values of $H$, as described by Eq.~(\ref{xquad1}). Right panel: the optimal realization of the fGn, $p(t)\equiv \dot{x}(t) +x(t)$ for the same set of $H$.}
  \label{fig1}
\end{figure}

Several examples of the optimal paths $x(t)$ for and $k=1$ and different $H$  are shown in the left panel of Fig. \ref{fig1}. The right panel shows, for the same values of $H$,  the optimal realizations of the fGn $\xi(t)$ itself, which are described by the expression $p(t)\equiv \dot{x}(t) + k x(t)$, see Eq.~(\ref{Langevin1}). Note that the path $X\to 0$, which brings the particle back to $x=0$, involves a nonzero noise. For $H$ not close to $1/2$, this noise is quite significant. (To remind the reader, for $H=1/2$ the ``relaxation path" is zero-noise, that is deterministic.) Overall, like in other systems, the knowledge of the optimal paths provides a valuable intuition about the physical mechanism behind the large deviations in question.

The reader may have noticed that the non-equilibrium optimal paths~(\ref{xquad1}) still exhibit a perfect mirror symmetry in time, $x(-t)=x(t)$, for all values of $H$. This is just a consequence of the time-reversibility
of the fGn and the linearity of the Langevin equation (\ref{Langevin1}) for $m=2$.

\subsubsection{$m=2$: autocovariance of the stationary fOU process}
\label{steady}

The optimal paths, dominating the single-point statistics for permissible $m$ and all $0<H<1$, are defined on the whole time axis, $|x|<\infty$, and they can be also found by assuming that the process $x(t)$ is \emph{stationary}. In the particular case of $m=2$, this stationary process -- the fOU process -- is also Gaussian. As a result, its one-point statistics can be found by directly minimizing the Gaussian action of the  process conditional on the value $X$. The corresponding Gaussian action is
\begin{equation}\label{stationarys}
S[x(t)] =  \frac{1}{2} \int_{-\infty}^{\infty} dt_1 \int_{-\infty}^{\infty} dt_2 K(t_1-t_2) x(t_1) x(t_2)\,,
\end{equation}
where $K(\tau)$ is the inverse kernel for the autocovariance $\kappa(\tau)$ of the stationary fOU process.  Recasting the condition $x(t=0)=X$ as an integral constraint and introducing a Lagrange multiplier $\lambda$ in the same way as in Eq.~(\ref{modified}), we can reduce the minimization problem to solving a simple integral equation:
\begin{equation}\label{stationaryeq}
\int_{-\infty}^{\infty} dt_1 K(t-t_1) \kappa(t_1) = \lambda \delta(t)\,.
\end{equation}
Its solution is
\begin{equation}\label{stationarysol}
x(t)=\lambda \kappa(t)\,.
\end{equation}
That is, the optimal path of the one-point statistics of the stationary fOU process coincides, up to a constant factor, with its autocovariance. An explicit formula for the latter  is obtained by dividing $x(t)$ from Eq.~(\ref{xquad1}) by $\lambda$ given by Eq.~(\ref{lambdavsX}), and the result coincides with the known expression \cite{Cheredito2003}.

In fact, this important connection between the optimal path of the one-point statistics and the autocovariance is more general, as it holds for \emph{all} stationary Gaussian processes \cite{Meerson2022,MO2022}.

\subsection{Asymptotic limits of $H\to 0$ and $H \to 1$}
\label{extremes}

\subsubsection{$H\to 0$}
\label{H0}

Sending $H$ to $0$ in Eq.~(\ref{P160}), we obtain $S(X,H=0) =\alpha(0,m) X^2/D$. Remarkably, the deterministic relaxation rate $k$ drops out. This implies that, in this limit, the action is entirely
determined by the noise and is independent of the external potential $V(x)$. In particular, the coefficient $\alpha(0,m)$ must be independent of $m$. Plugging $H=0$ into  Eq.~(\ref{alpham=2}), we obtain $\alpha(0,2)=1/2$. Therefore, $\alpha(0,m)=1/2$ for all permissible $m$, and
we arrive at a universal (potential-independent) $H\to 0$ asymptotic of the action:
\begin{equation}\label{PP020}
S_{H\to 0}(X)=\frac{X^2}{2 D}\,.
\end{equation}
In the frequency domain, this asymptotic is dominated by the high frequencies of the fGn, $|\omega| \gg k$.
For $m=2$ this fact becomes evident. Indeed, the calculation of the action by using Eq.~(\ref{AA030})
is reduced, at any $H$, to evaluating the integral
\begin{equation}\label{integral2}
\int_{-\infty}^\infty d\omega\,\frac{|\omega|^{1-2H}}{\omega^2+k^2}\,.
\end{equation}
At $H\to 0$, this integral is dominated by large $\omega$, up to $|\omega|\sim k \exp[1/(2H)]$.

\subsubsection{$H\to 1$}

When $H$ approaches $1$ in Eq.~(\ref{P160}), we obtain $S(X,H\to1) =\alpha(1,m) f(X)^2/D$.
As we will see, this will lead us to the conclusion that $\alpha(1,m)$ does not depend on $m$ as well. Let us consider the particular case of $m=2$. When $H\to 1$, Eq.~(\ref{alpham=2})  becomes
\begin{equation}\label{PP030}
S_{H\to 1}(X)=\frac{f(X)^2}{4 D}=\frac{(kX)^2}{4 D}\,.
\end{equation}
In this limit the integral in Eq.~(\ref{integral2}) is mostly contributed to by the low frequencies,
$|\omega| \ll k$. Here one can neglect the term $\dot{x}$ in the Langevin equation~(\ref{Langevin1}). Therefore, to zero order, we have a balance $-kx(t)\simeq \bar{\xi}_T(t)$ at $T\gg k^{-1}$,
where
\begin{equation}\label{lowfreq}
\bar{\xi}_T(t)=\frac{1}{2T}\int_{-T}^{T} \xi(t+\tau)\, d\tau\,,
\end{equation}
is the low-frequency part of the fGn $\xi$. Using Eq.~(\ref{corry}) for $\langle\xi(t)\xi(t^\prime)\rangle$, we obtain at a given $T$ and $H$ tending to 1 that
$\langle\bar{\xi}_T^2\rangle\bigr|_{H\to 1}=2D$. Therefore, Eq.~(\ref{PP030}) describes the Gaussian distribution $\mathcal{P}(\bar{\xi}_T)$ of  $\bar{\xi}_T$,  that is,
\begin{equation}\label{PP040}
\mathcal{P}_{H\to 1}(\bar{\xi}_T)\simeq \frac{1}{\sqrt{4\pi D}}\exp \left(-\frac{\bar{\xi}_T^2}{4D}\right)\, .
\end{equation}
Similarly, the approximate force balance $-f(x) \simeq \xi_T(t)$ holds,  in the limit of $H\to 1$,  for all monotonically growing forces $f(x)$. Using  the force balance and Eq.~(\ref{PP040}), we obtain the PDF of $X$ we are after:
\begin{equation}\label{PP059}
\mathcal{P}_{H\to 1}[f(X)]\simeq \frac{1}{\sqrt{4\pi D}}\exp \left[-\frac{f(X)^2}{4D}\right]\,.
\end{equation}
In particular, for the power-law potentials~(\ref{f(x)}) with $m>1$, Eq.~(\ref{PP059}) becomes
\begin{equation}\label{PP060}
\mathcal{P}_{H\to 1}(X)\simeq \frac{k(m-1)|X|^{m-2}}{\sqrt{4\pi D}}\exp \left(-\frac{k^2|X|^{2m-2}}{4D}\right)\,,
\end{equation}
As one can see, in this limit  we have been able to account for the pre-exponential factor of the distribution as well, and therefore to describe the whole  distribution~(\ref{PP060}). Strikingly, for $m>2$ the distribution (\ref{PP060}) is bimodal: it vanishes at $X=0$.  The bimodality of $\mathcal{P}(X)$ at $m>2$ in the particular limit of $H\to 1$ agrees  with the findings of Ref. \cite{Metzler2021} who observed bimodality numerically for all $H>1/2$.

The action, corresponding to Eq.~(\ref{PP060}), is
\begin{equation}\label{PP070}
S_{H\to 1}(X)\simeq \frac{k^2|X|^{2m-2}}{4D}\,.
\end{equation}
Comparing this expression with Eq.~(\ref{P160}), we conclude that $\alpha(1,m) = 1/4$ independently of $m$.

Altogether, we know the values of the factor $\alpha(H,m)$ in Eq.~(\ref{P160}) in three limits:
\begin{equation}
\alpha(H,m) =
    \begin{cases}
         \frac{1}{2}\,, & H\to 0\,,\\
         \frac{1}{m}\,, & H=1/2\,,\\
         \frac{1}{4}\,, & H\to 1\,,
    \end{cases}
\label{threelimits}
\end{equation}
where the first and third lines do not depend on $m$.

\subsection{Path irreversibility}
\label{violation}

One important signature of nonequilibrium is path irreversibility: a disparity between the back and forth probabilities of the system moving along a closed path in the parameter space. We can use the expression~(\ref{Ssub}) to calculate the difference $S[x(t)]-S[x(-t)]$:
\begin{equation}\label{irr}
S[x(t)]-S[x(-t)]=-2\int\limits_{-\infty}^\infty\int\limits_{-\infty}^\infty C(t-t^\prime)\dot{x}(t)f(x(t^\prime))\,.
\end{equation}
This expression vanishes for $H=1/2$, where closed paths, obeying~(\ref{BCs}), are reversible for any $m>0$. For $H\neq 1/2$ they are irreversible except for
$m=2$ when the Langevin equation~(\ref{Langevin1}) is linear.

\begin{figure}[ht]
  %\centering
  \includegraphics[width=8.0cm]{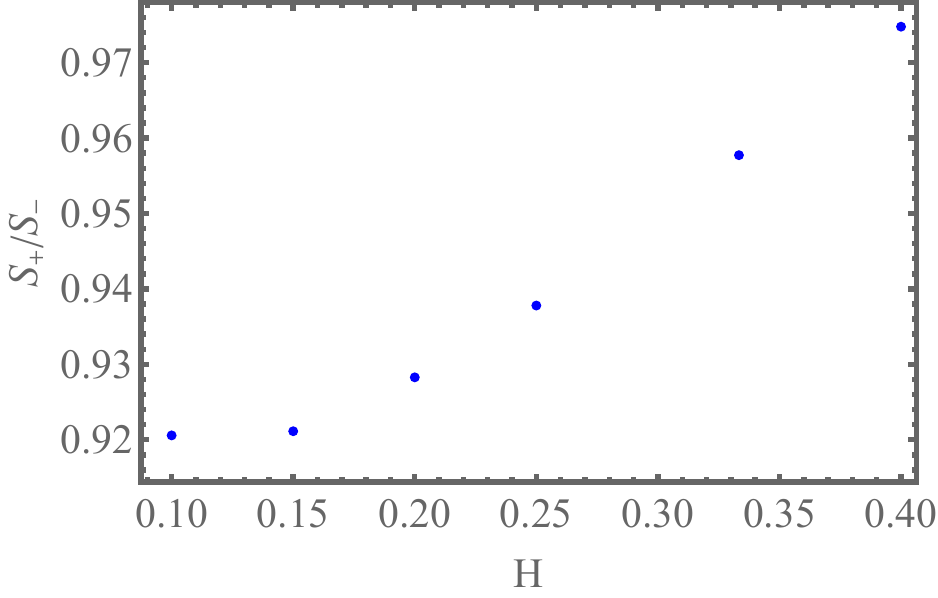}
  \caption{The ratio of the forward and backward actions~(\ref{Ssub}) vs. $H$ for the closed path~(\ref{closedpath}). The parameters are $D=1/2$ and $k=X=1$.}
  \label{ratioS}
\end{figure}

As an illustrative example of irreversibility, let us consider an asymmetric closed path $0\to X \to 0$:
\begin{equation}\label{closedpath}
  x(t) =
    \begin{cases}
         X\,e^t\,, & t<0\,,\\
        X\,e^{-2t}\,, & t>0\,\\
    \end{cases}
\end{equation}
in a quartic potential, $m=4$. We evaluated the action~(\ref{Ssub}) for several values of $0<H<1/2$ by performing the integrations over time  from $t=-\infty$ to $\infty$ and backward. Let us denote the resulting forward and backward actions by $S_+(H)$ and $S_-(H)$, respectively.
Figure \ref{ratioS} shows the ratio $S_+/S_-$ as a function of $H$, and irreversibility is evident.
As $H$ approaches $1/2$, the ratio $S_+/S_-$ approaches $1$ as to be expected.

\section{Regularization of Eq.~(\ref{inteq1}) and numerical optimal paths}
\label{numerics}

\subsection{Regularization}
\label{reg}

Throughout this section we confine ourselves to the subdiffusion case, $0<H<1/2$. Inspecting the $t\to 0$ asymptotic behavior of the exact optimal path $x(t)$  for $m=2$, Eq. (\ref{smallt}), and the relationship~(\ref{lambdavsX}) between $\lambda$  and $X$, one can notice that they are independent of $m$. Therefore,  the character of the cusp singularity of $x(t)$ at $t\to0$,
\begin{equation}\label{R010}
 x(t)\bigr|_{t\to0}=X-\lambda D |t|^{2H}+\dots\,,
\end{equation}
remains the same for all $m>0$. This reflects the fact that, for $H<1/2$, the second term in the asymptotic~(\ref{R010}) is universal and independent of $m$ (actually, of the confining force $f(x)$ in general). This observation plays a critical role in our approach to a numerical solution of Eq.~(\ref{inteq1}).

The most singular term in the left hand side of Eq.~(\ref{inteq1}) comes from the second derivative of the second term in Eq.~(\ref{R010}). A further observation is that the first two terms of the asymptotic~(\ref{R010}) provide an exact solution of a \emph{truncated} Eq.~(\ref{inteq1}) which includes only the most singular term on the left hand side -- the term originating from $\ddot{x}$) -- and the delta-function term on the right hand side.  All the other contributions, which come from the rest of the left hand side terms and from the higher-order terms in the expansion~(\ref{R010}), are much less singular at $t\to 0$.

As one check, the second term in the expansion~(\ref{R010}) is so singular that it makes it impossible to treat the integral in Eq.~(\ref{inteq1}) as an improper integral at $t^\prime=0$ of a function that is regular elsewhere.
As a result, Eq.~(\ref{inteq1}) cannot be used for numerical analysis in its present form. The main goal of this section is to obtain a regularized form of Eq.~(\ref{inteq1}), and we achieve this goal via an analytical treatment of the most singular part of $x(t)$ coming from the 2nd term in the expansion~(\ref{R010}). To this end we present $x(t)$, for all $t$, in the following form:
\begin{equation}\label{R020}
 x(t)=\tilde{x}(t)-\lambda D |t|^{2H}\,,
\end{equation}
where $\tilde{x}$ is a much smoother function at $t=0$ than $x(t)$, and try to derive a regular equation for $\tilde{x}$.

For brevity, we rewrite Eq.~(\ref{inteq1}) in the following form:
\begin{equation}\label{NN360}
 C*\ddot{x} +f^\prime(x)\cdot C*\dot{x}-C*f^\prime(x)\cdot \dot{x}
  - f^\prime(x)\cdot C*f(x)+\lambda\delta(t)=0\, .
\end{equation}
Here all functions are functions of $t$, $C(t)$ is defined by Eq.~(\ref{solC2}), the symbol
$F\cdot$ denotes ordinary multiplication by the function $F$, and the asterisk denotes the convolution of two functions of $t$. Rearranging the terms in this equation and using the identity $c(t)*C(t)=\delta(t)$, we can rewrite Eq.~(\ref{NN360}) in an equivalent form,
\begin{equation}\label{NN368}
 C*\left[\ddot{x}-f(x)f^\prime(x)+ \lambda  c(t)\right]
=\left[ f^\prime(x)\cdot,\, C\, *\right]\left(f-\dot{x}\right)\,,
\end{equation}
where the commutator $\left[ f^\prime(x)\cdot,\,\hat{C}\, *\right]$ is defined as
\begin{equation}\label{NN380}
\left[ f^\prime(x)\cdot,\, C\, *\right]\, F= f^\prime(x)\cdot C *F- C *f^\prime(x)\cdot F\,.
\end{equation}
[Note that the commutator in Eq.~(\ref{NN368}) vanishes at $m=2$ for any $H$, or at $H=1/2$ for any $f(x)$.] Making the convolution of the both sides of Eq.~(\ref{NN368}) with the kernel $c(t)$, we obtain
\begin{equation}\label{NN370}
\ddot{x}-f(x)f^\prime(x)+ \lambda c(t)
= c*\left[ f^\prime(x)\cdot,\,C\, *\right]\left(f(x)-\dot{x}\right)\,.
\end{equation}
Using Eqs.~(\ref{corry}) and~(\ref{solC2}) for $c(t)$ and $C(t)$, respectively, we can rewrite Eq.~(\ref{NN370}) as
\begin{equation}\label{NN560}
\frac{d^2}{dt^2}\left\{x(t)+D\lambda |t|^{2H} +Z[x](t)\right\}
   = f(x(t))f^\prime(x(t))\,,
\end{equation}
where
\begin{equation} \label{NN550}
 Z[x](t) =C_0\,|t|^{2H}\left[ f^\prime(x(t))\cdot,\,|t|^{-2H}\, *\right]\left[\dot{x}(t)-f(x(t))\right]
\end{equation}
and $C_0= \cot (\pi H)/(4\pi H)$. Going over from the dependent variable $x(t)$ to the new dependent variable
\begin{equation}\label{NN580}
\bar{x}(t)=x(t)+D\lambda |t|^{2H} +Z[x](t)\,,
\end{equation}
we arrive at the following equation for $\bar{x}(t)$:
\begin{equation}\label{NN590}
\frac{d^2}{dt^2}\bar{x}(t)=f(x)f^\prime(x)\,,
\end{equation}
where the old variable $x(t)$ in the right-hand-side should be expressed through $\bar{x}(t)$:
\begin{equation}\label{xviaxbar}
x(t)=\bar{x}(t)-\lambda D|t|^{2H}-Z[x](t) =\tilde{x}(t)-Z[x](t)\,.
\end{equation}
Note that $Z[x](t)$ vanishes identically  at $m=2$ for any $H$ and at $H=1/2$ for any $m$. In these cases Eq.~(\ref{NN590}) turns into a second-order ODE (linear or nonlinear, respectively) which can be exactly solved for $\bar{x}(t)$.
As to be expected, the resulting $x(t) = \bar{x}(t)-\lambda D|t|^{2H}$ perfectly agrees with the corresponding solutions in these cases that we presented above.

As we  argue below, in the general case $Z[x](t)$ is a continuous function of $t$ at least on the solutions of Eq.~(\ref{NN560}) [or~(\ref{NN590})]. Let us assume for a moment that this continuous function $Z[x](t)$ is known. Then Eq.~(\ref{NN590}) becomes an ordinary differential equation (ODE), and it has solutions $\bar{x}(t)$ which are twice differentiable. Some examples of iterative numerical solutions of the problem, defined by Eqs.~(\ref{NN590}), (\ref{xviaxbar}) and (\ref{BCs}), will be presented in Sec.~\ref{numnum}.

Rigorously speaking, the integral corresponding to the second convolution, $|t|^{2H}*\dots$, in Eq.~(\ref{NN550}) for $Z[x](t)$, diverges at  $t^\prime\to\pm\infty$. However, using the freedom of adding to $Z[x](t)$ any linear function of $t$ without affecting the ODE, we can ``improve" the definition of $Z$ so that it becomes bounded for all finite times, with the same solution $x(t)$. Such an improved definition is considered in Appendix~\ref{ap2}. This redefinition, however, becomes unnecessary in the numerical solution of Eq.~(\ref{NN560}) due to the finiteness of the numerical time interval, see Appendix~\ref{ap1}.

Now we present our argument that $Z[x](t)$ is a continuous function of $t$  at least on the solutions of Eq.~(\ref{NN560}). Let us recast the definition of $Z[x](t)$ in Eq.~(\ref{NN550}) in a shorter form,
\begin{equation}\label{NN591}
Z[x](t)\propto |t|^{2H}*Y[x](t)\,,
\end{equation}
where
\begin{equation}\label{NN592}
Y[x](t)=\left[ f^\prime(x(t))\cdot,\,|t|^{-2H}\, *\right]\left[\dot{x}(t)-f(x(t))\right]\, .
\end{equation}
The most singular term at $t=0$ in Eq.~(\ref{NN592}) comes from the derivative of the second term in the asymptotic~(\ref{R010}). Keeping only this singular term in Eq.~(\ref{NN592}), we see that the most singular part of the expression for $Y$ at $t\to 0$  is the following:
\begin{equation}\label{NN594}
Y[x](t)_{t\to0} \propto \mbox{p.v.}\int \frac{|t^\prime|^{2H} \left(|t^\prime|^{2H}-|t|^{2H} +\dots \right)\, dt^\prime}{t^\prime\,  |t-t^\prime|^{2H}} +\dots \,,
\end{equation}
where  we have kept only the most singular terms in the vicinity of $t^\prime=0$ when $t\to 0$. Now we can see that all the singularities of the integrand are integrable at  $t^\prime = 0$ and $t^\prime=t$.
As a result, we have:
\begin{equation}\label{Yt0}
Y[x](t)_{t\to0}=A+B|t|^{2H}+\dots\,,
\end{equation}
with some constant $A$ and $B$. It is evident now that the leading  singularity of $Y$ at $t\to 0$ is the same as the singularity of $x(t)$, and that the function $Y[x](t)$ is continuous at $t=0$. Therefore, by virtue of Eq.~(\ref{NN591}), $Z[x](t)$ is also a continuous function of $t$ . Our numerical solutions of this problem, presented in Sec.~\ref{numnum}, confirm these (not entirely rigorous) arguments.

\subsection{Numerical solution}
\label{numnum}

In the general case of a nonzero $Z[x](t)$, Eq.~(\ref{NN560}) is a nonlinear integro-differential equation which cannot be solved analytically. Here we propose a numerical iteration procedure which allows one to get sufficiently accurate numerical solutions
of Eq.~(\ref{NN560}), and hence Eq.~(\ref{inteq1}), for $H<1/2$ in spite of the expected singularity of the solution at $t=0$.

The numerical iteration procedure treats  Eq.~(\ref{NN590})  as a second-order ODE for $\bar{x}(t)$, defined by Eq.~(\ref{NN580}), with the boundary conditions
\begin{equation}\label{xbarBCs}
x(t=-T) = x (t=T) = 0\,,
\end{equation}
where $T$ is sufficiently large but finite.  To obey the boundary conditions  we used a shooting method which involves a numerical  solution of the Cauchy problem with the fixed condition $x(t=-T)=0$ but an adjustable condition for $\dot{\bar{x}}(t=-T)$. The function $Z[x](t)$ is calculated by using $\tilde{x}(t)$ from the previous iteration. We found out that this iteration procedure converges quite fast to its limiting solution. Some additional details of the iteration procedure are presented in Appendix~\ref{ap1}.

We used this iteration procedure to compute the optimal paths $x(t)$ and the actions for  $m=5/2$ and three values of  $H$: $1/8$, $1/4$ and $3/8$. We set  $k=1$ and $\lambda D=1/2$ without limiting generality.  Having found numerically $x(t=0)\equiv X$, we computed the dimensionless coefficient $\alpha(H,m)$ for the chosen values of $H$ and $m$ by substituting  Eq.~(\ref{P160}) for the action $S$ into the shortcut relation~(\ref{shortcut}):
\begin{equation}\label{shortcutnum}
\frac{dS}{dX}\biggr|_{X=x(0)}=\frac{1}{2D}\,.
\end{equation}
This gives
\begin{equation}\label{alphax(0)}
\alpha(H,m)=\frac{X^{4H-1-2Hm}}{4(1-2H+Hm)}\, .
\end{equation}
The coefficients $\alpha(H,m)$ in Eq.~(\ref{P160}), computed in this way, are presented in Fig.~\ref{figa}.

\begin{figure}[ht]
  %\centering
  \includegraphics[width=0.45\textwidth]{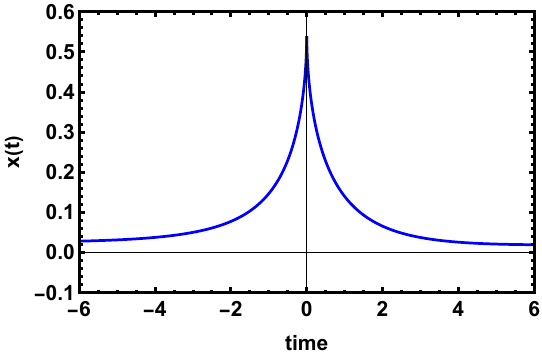}
  \caption{Numerical optimal path $x(t)$ for  $m=5/2$ and  $H=1/4$. The parameters $k$ amd $\lambda D$ are  $k=1$ and $\lambda D=1/2$.}
  \label{fig3}
\end{figure}

\begin{figure}[ht]
  %\centering
  \includegraphics[width=0.45\textwidth]{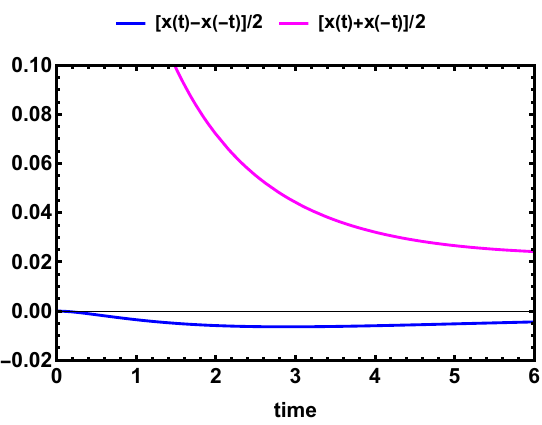}
  \caption{Irreversibility of the optimal path shown in Fig.~\ref{fig3}. Plotted are the functions  $(x(t)-x(-t))/2$ and $(x(t)+x(-t))/2$.}
  \label{fig4}
\end{figure}

To avoid similarly looking figures for the optimal paths, here we present the numerically found optimal path $x(t)$ only for $H=1/4$, see Fig.~\ref{fig3}. At $t=0$ $x(t)$ has a cusp singularity of the same type as the one for $m=2$, see Fig.~\ref{fig1}. This supports our argument presented in Sec.~\ref{reg}.  As explained in Sec.~\ref{violation}, the optimal path is expected to show irreversibility, that is to be asymmetric with respect to the reflection $t\leftrightarrow -t$. The asymmetry is not so evident in Fig.~\ref{fig3}, but it becomes evident in the plots of the  functions $(x(t)-x(-t))/2$ and $(x(t)+x(-t))/2$, see Fig.~\ref{fig4}.

\begin{figure}[ht]
  %\centering
  \includegraphics[width=0.45\textwidth]{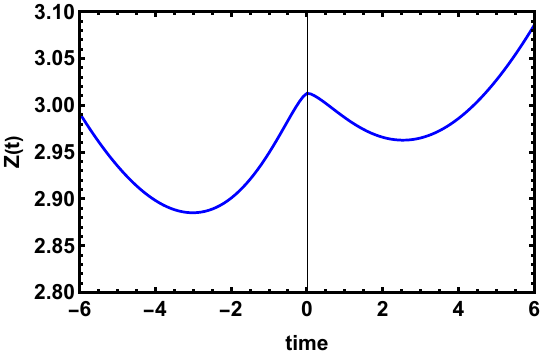}
  \caption{The function $Z[x](t)$ evaluated on  the numerical optimal path $x(t)$ for $H=1/4$, shown in Fig.~\ref{fig3}.
  }
  \label{fig5}
\end{figure}

Figure~\ref{fig5} shows $Z(t)$ computed by using the numerically found $x(t)$ shown in Fig.~\ref{fig3}. The function $Z(t)$ is manifestly continuous, in line with our argumentation in Sec.~\ref{reg}. We also checked that the continuity of $Z[x](t)$ is preserved at each step of the iteration procedure.

\section{Summary and Discussion}
\label{discussion}

As we have demonstrated, the OFM provides a valuable insight into the (in general, nonequilibrium) steady state of fBM confined by a scale-invariant power-law potential. The OFM describes the tails of the steady-state distribution and gives, for arbitrary $H$ and $m$, exact conditions for confinement. In the particular case of a quadratic potential, which corresponds to
the fOU process, the OFM provides exact results not only for the whole steady-state probability distribution of
the particle position, but also for the autocovariance of the stationary process.  Additional analytical results can be obtained in the asymptotic limits of $H\to0$ and $H\to1$.
Importantly, these asymptotic results apply to a broad class of confining potentials, not necessarily scale-invariant. For non-quadratic potentials, the OFM problem for the position distribution can be solved numerically.

In all cases, the OFM predicts, analytically or numerically, the optimal paths of the process, conditioned on $X$. They highlight the large-deviation physics of this process. For example,  one can contrast the cusps and  the non-monotonic time dependence of $x(t)$ for the antipersistent fBm ($H<1/2$) and the smooth fat-tailed  $x(t)$ for the persistent fBm ($H>1/2$).

Here is an additional useful insight from the optimal paths. The form of Eqs.~(\ref{Ssub}) and~(\ref{solC2}) might lead one to assume that
the action $S(H)$  should  behave like $1/H^2$ at $H\to 0$, reflecting the $1/H^2$ singularity of the inverse kernel $C(\tau)$.  Similarly, one may assume that $S(H)$ behaves as $1/(1-H)$ at $H\to 1$, see Eqs.~(\ref{actionforcesuper}) and (\ref{kernelmore}).
Such singularities would indeed occur if the optimal paths were smooth and had a characteristic time scale on the order of $1$, dictated by the potential $V(x)$ alone. It is clear from our results, however, that the actual optimal paths are quite different. As a consequence, $S(H)$ does not exhibit any singularity for all $0\leq H\leq 1$. It is quite likely that similar effects can be found in additional systems driven by a long-correlated noise.

It would be very interesting to test our predictions for the tails of $\mathcal{P}(X)$, and for the corresponding optimal paths, in large-deviation simulations of the fractional Langevin equation (\ref{Langevin1}). Very recently, large-deviation simulations of a free (unconfined) fBm  have been successfully accomplished in the context of the first-passage area distribution and its optimal paths \cite{HM2024}.

One natural extension of the OFM for the fBm in a confining potential concerns the \emph{first passage}  problem
\cite{Metzler2010,Wang2017}. For the scale-invariant confining potentials, and  under condition that the first passage is a rare event, our Eq.~(\ref{P160}) for the action $S(X)$ also determines, up to a preexponential factor, the mean first-passage time $\langle T\rangle \sim \exp[S(X)]$ to a specified point $x=X$. In their turn, the optimal paths $x(t)$, that we studied in this work, are also the optimal paths of the first passage to the point $X$. In particular, our explicit analytical results for $m=2$ are directly relevant to the small-$D$ regime of  the first-passage setting studied numerically in Ref. \cite{Metzler2010}.

\section*{Acknowledgments}

We are very grateful to Alexander K. Hartmann and Alexander Valov for useful discussions. B. M. was supported by the Israel Science Foundation (Grant No. 1499/20).

\vspace{0.5 cm}

\appendix
\section{Redefinition of $Z[x](t)$}
\label{ap2}

Let us recast the definition~(\ref{NN550}) of $Z[x](t)$ in the following form:
\begin{equation}\label{NN610}
Z[x](t)
=C_0\,\int\limits_{-\infty}^\infty dt^\prime\,|t-t^\prime|^{2H} Y[x](t^\prime)\, ,
\end{equation}
where
\begin{equation}\label{NN620}
Y[x](t)=\left[ f^\prime(x(t))\cdot,\,|t|^{-2H}\, *\right]\left[\dot{x}(t)-f(x(t))\right]\, .
\end{equation}
The integral in Eq.~(\ref{NN610}) over $t^\prime$ diverges at $t^\prime\to\pm\infty$ for any $t$. To cure this divergence, we can improve the definition~(\ref{NN610}) by introducing
\begin{equation}\label{NN640}
Z_T[x](t)
=C_0\,\int\limits_{-T}^T dt^\prime\,|t-t^\prime|^{2H} Y[x](t^\prime)
\end{equation}
and by defining $Z[x](t)$ as follows:
\begin{equation}\label{NN650}
Z[x](t)=\lim\limits_{T\to\infty} \biggl\{Z_T[x](t)-Z_T[x](0)-t\frac{Z_T[x](T_1)-Z_T[x](-T_1)}{2T_1}\biggr\}\, ,
\end{equation}
where $T_1$ is an arbitrary positive constant of order $1$.

\section{Details of the iteration procedure}
\label{ap1}

We implemented the iteration procedure, outlined in Sec.~\ref{numnum} for the function $\tilde{x}(t)$, which is defined as
\begin{equation}\label{NN434}
\tilde{x}(t)=x(t)+\lambda D|t|^{2H}
\end{equation}
and presumably has a continuous first derivative. We consider this function only on the interval $|t|<T$, where $0<T<\infty$. As a result, all the integrals, entering the definition of the convolutions in Eq.~(\ref{NN550}), are computed on the same time interval. For this reason there is no need for a redefinition of $Z[x](t)$, caused by the formal divergence of the second convolution in the expression~(\ref{NN550}) at $T=\infty$.

During the iterations of
$\tilde{x}_n$,
\begin{equation}\label{NN433}
\tilde{x}_n\to \tilde{x}_{n+1}\quad (n=0,1,2,\dots)\,,
\end{equation}
the terms $x_n(t)$, which appear below, are evaluated by using Eq.~(\ref{NN434}) for the known $\tilde{x}_n(t)$. The derivative of $x_n(t)$ is evaluated by computing the numerical derivative of $\tilde{x}_n$ and the explicit analytical derivative of the second term in Eq.~(\ref{NN434}). Each  step of the iterations consists of several sub-steps which are listed below. The expression
\begin{eqnarray}
  Y[\tilde{x}_n](t) &=& \int\limits_{-T}^T  dt^\prime\, |t-t^\prime|^{-2H}\left[f^\prime(x_n(t))-f^\prime(x_n(t^\prime))\right]\left[\dot{x}_n(t^{\prime})
-f(x_n(t^{\prime}))\right] \nonumber\\
  &=& \int\limits_{-T}^T  dt^\prime\, |t-t^\prime|^{-2H}\left[f^\prime(x_n(t))-f^\prime(x_n(t^\prime))\right]\left[\dot{\tilde{x}}_n(t^{\prime}) -2\lambda D H |t^\prime|^{2H-1}\, \mbox{sign}\, t^\prime -f(x_n(t^{\prime}))\right]
  \label{NN432}
\end{eqnarray}
evaluates $\left[ f^\prime(x(t))\cdot,\,|t|^{-2H}\, *\right]\left[\dot{x}(t)-f(x(t))\right]$ in Eq.~(\ref{NN550}). Further,
\begin{equation}\label{NN436}
Z[\tilde{x}_n](t)=C_0\,\int\limits_{-T}^T dt^{\prime}\,
|t-t^{\prime}|^{2H} Y[\tilde{x}_n](t^\prime)\,.
\end{equation}
We introduce
\begin{equation}\label{NN438}
X[\tilde{x}_n](t)=\tilde{X}[\tilde{x}_n](t)-\lambda D|t|^{2H}\,;
\quad
\tilde{X}[\tilde{x}_n](t)=\bar{X}[\tilde{x}_n](t)-Z[\tilde{x}_n](t)\,.
\end{equation}
Now we solve numerically the ODE
\begin{equation}\label{NN430}
\ddot{\bar{X}}-f(X)f^\prime(X)=0
\qquad
\left(X(t)_{|t|= T}= 0\right)\, .
\end{equation}
Then we calculate $\tilde{x}_n$ of the next step of the iterations as follows:
\begin{equation}\label{NN440}
\tilde{x}_{n+1}(t)=(1-\alpha_n)\, \tilde{x}_n(t)+\alpha_n \tilde{X}[\tilde{x}_n](t)\, .
\end{equation}
$\alpha_n=1$ would correspond to the straightforward iterations briefly considered in Sec.~\ref{reg}: $\tilde{x}_{n+1}(t)=\tilde{X}[\tilde{x}_n](t)$. However, to avoid a possible instability of the iteration process we used the more conservative weights
$$
\alpha_0=1\quad \mbox{and}\quad\alpha_{n\geq 1}=\alpha=0.5\,.
$$
We started the iteration process by settting $x_0(t)=0$, so that $Z[x_0](t)=0$, and $\bar{x}_1=\tilde{x}_1=\tilde{X}=\bar{T}$ is the solution of the ODE $\ddot{\bar{X}}-f(X)f^\prime(X)\bigr|_{X=\bar{X}-\lambda D|t|^{2H}}=0$.

We observed, for $H=1/4$ and $m=5/2$,  that replacing $T=\infty$ by $T=15$ does not lead to an error in the value of $x(0)$ that would exceed $\pm0.5\%$. The iterations converge relatively fast: about 10-15 iterations were sufficient to find the limiting optimal path for $T=15$ within the above-mentioned inaccuracy of less than $\pm 0.5\%$ caused by the finite $T=15$.

\end{document}